\documentclass[11pt,a4paper]{article}

\renewcommand{\author}{E.\ Khalisi}
\newcommand{\titel}{Did a Series of Solar Eclipses Inspire the Nazca Lines?}
\newcommand{\version}{Version 1.92}
\renewcommand{\date}{\today}

\usepackage{hyperref}
\hypersetup{pdfauthor={Emil Khalisi}}
\hypersetup{pdftitle={\titel }}
\hypersetup{pdfsubject={Online publication, \version\ of \date\ }}
\hypersetup{pdfkeywords={Eclipses, Archaeoastronomy, Geoglyphs, Nazca lines, Peru}}
\hypersetup{colorlinks=true,citecolor=blue}

\usepackage[T1]{fontenc}        
\usepackage{mathptmx}           
\usepackage{pifont}             
\usepackage[scaled=0.92]{helvet} 

\usepackage[british]{babel}     
\usepackage{amssymb}            
\usepackage{microtype}          

\usepackage{multicol}           
\usepackage{multirow}           
\usepackage{arydshln}           
\usepackage{paralist}           
\usepackage{units}              

\usepackage{calc}
\usepackage[a4paper]{geometry}  
\usepackage{scrextend}           
\changefontsizes{10pt}
\usepackage{titlesec}
\titleformat*{\section}{\large\bfseries}
\titleformat*{\subsection}{\normalsize\bfseries}

\usepackage{graphicx}           
\usepackage{float}              
\usepackage{caption}            
\usepackage{fancyhdr}
\renewcommand{\headrulewidth}{0.4pt}
\usepackage{colortbl}             
\definecolor{grey20}{RGB}{208,208,208}
\definecolor{grey10}{RGB}{230,230,230}

\usepackage{url}
\usepackage[numbers]{natbib}     
\begin{document}


\fancyhead{}
\fancyhead[LO]{%
   \footnotesize \textsc{In original form published in:}\\
   {\footnotesize Habilitation at the University of Heidelberg }
}
\fancyhead[RO]{
   \footnotesize {\tt arXiv:(side bar) [physics.hist-ph]}\\
   \footnotesize {Date: 26 March 2021}%
}
\fancyfoot[C]{\thepage}

\renewcommand{\abstractname}{}

\twocolumn[
\begin{@twocolumnfalse}

\section*{\centerline{\LARGE \titel }}

\vspace{\baselineskip}
\begin{center}
{Emil Khalisi\\}
\textit{D--69126 Heidelberg, Germany}\\
\textit{e-mail:} \texttt{ekhalisi[at]khalisi[dot]com}\\
%
\end{center}


\vspace{-\baselineskip}
\begin{abstract}
\changefontsizes{10pt}
\noindent
\textbf{Abstract.}
The mysterious lines in the desert of Nazca in southern Peru received
many attempts at an explanation.
We discuss the possibility that their origin is based on solar
eclipses which could have evoked some kind of ``reaction'' by the
inhabitants.
In the mid-4th century BCE, six very conspicuous events within 20
years occurred over that area.
The climax was reached with the total eclipse of 11 March -357.
In this paper we try to explain the geoglyphs as a medium of
communication of the ancients with their gods.

\vspace{\baselineskip}
\noindent
\textbf{Keywords:}
Eclipses,
Archaeoastronomy,
Geoglyphs,
Nazca lines,
Peru.
\end{abstract}

\centerline{\rule{0.8\textwidth}{0.4pt}}
\vspace{2\baselineskip}

\end{@twocolumnfalse}
]




\section{Introduction}

The coastline of present-day Peru was once inhabited by a culture
named after the city of Nazca about 450 km south of Lima.
This people of the Andes evolved in pre-Inca times, probably at
about 200 BCE, under an extreme climate:
the region is one of the driest on Earth.

\begin{figure}[t]
\includegraphics[width=\linewidth]{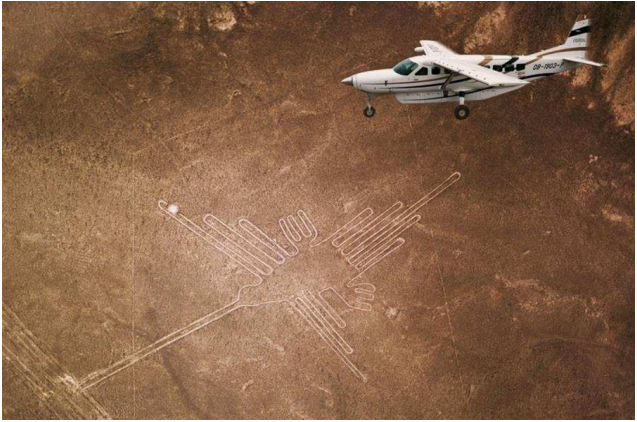}
\caption{Hummingbird as an example for the geoglyphs at Nazca in Peru
   \cite{fathy:2019}.}
\label{fig:nazcalines}
%
%
\end{figure}

It is believed that the Nazca had no centrally governed empire,
but were an agglomeration of tribes.
They have become famous for the geoglyphs (or ``earth drawings'')
shaping huge figures to be only seen from high elevation
(Fig.\ \ref{fig:nazcalines}).
The viewing position was inaccessible at times before aviation,
therefore, the figures were discovered by a pilot in 1927 by chance
when he spotted extended patterns arranged in the landscape.

There are hundreds of drawings classified into three main groups:
\begin{compactitem}
\item very long straight lines (some over 10 km),
\item geometric figures (trapeziums, spirals, zigzag lines),
\item and representations of living beings (birds, spiders, apes,
   etc., and a few flowers).
\end{compactitem}

Countless articles and many books were written by scientists and
non-professionals alike.
They deal with various issues comprising the manufacture of the
lines, surveys of their forms, statistics regarding orientation,
cultural relation, and, in particular, their purpose.
This last question is the most enigmatic and acts like an invitation
for some kind of contribution even to laymen.
Many minds tried their best to solve the secret that seems like
an uncrackable nut.
The different theories run up to as high as the number of
publications.

Here, we do not aim for a detailed analysis of the lines but rather
for more compelling reasons related to solar eclipses.
This possibility was brought up on the Internet and seems worth
discussing.
Our aspiration is to include this option to the variety of
suggestions.
We are fully aware that our explanation might be wrong, either.
However, we discovered a period of over two decades accumulating
more than ten eclipses of remarkably high magnitude that might have
influenced the people.

For those who have not heard about the Nazca lines, we briefly
summarise a little basics first.
Deeper insight can be found in the monumental study by Anthony Aveni
\cite{aveni:1990}.
Dates will be given astronomically, for example, the year ``-374''
will be ``375 BCE'' for the historians.


\section{Background information}

The geoglyphs of Nazca extend over 50 km$^2$ across the desert in
southern Peru.
They are extraordinary insofar as they were never sighted by their
creators in full magnificence.
A peculiarity is that the biomorphic figures are made up of one
continuous line that does not intersect itself.

The largest individual feature measures 370 metres in length,
and the longest of the straight lines is 14.5 km.
Taken all lines together the total length exceeds 1,300 km.
In 2019, over 140 new figures were discovered employing advanced
technology like drones to map them in unprecedented detail
\cite{eda:2019}.
Some of them scarcely contrast against the surrounding and are
difficult to detect.
Geoglyphs can be found on other places, too, e.g.\ in Egypt, Malta,
Bolivia or in the US,
but they are nowhere as numerous and stylish as in Nazca
\cite{nickell:2005}.

\begin{figure}[t]
\includegraphics[width=\linewidth]{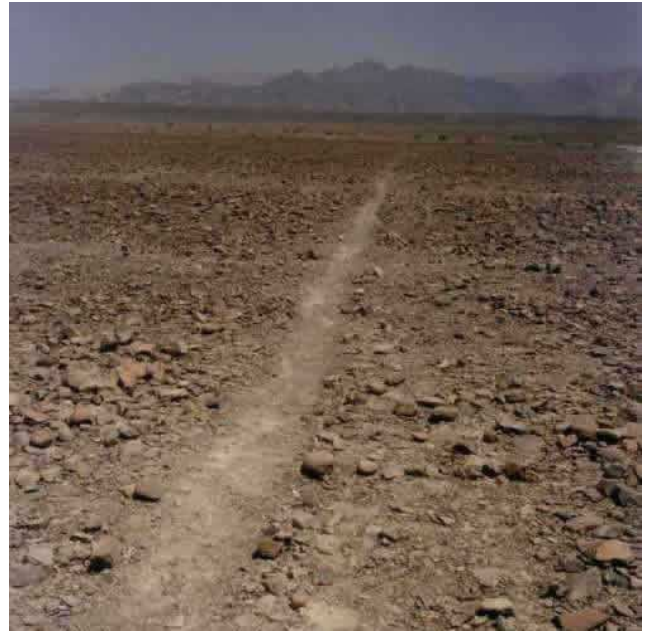}
\caption{In a close-up view the line appears like a pathway.}
\label{fig:nazcapath}
\end{figure}

The time of their origin is not well known.
It is thought that they were etched into the desert ground some
time between 200 BCE and CE 600.
This estimate is based on radiocarbon dating of organic material
forming ``rock varnish'' \cite{dorn:1991}.
The soil is not sandy but hard-packed consisting of dark, iron
oxide-coated gravel.
The producers removed the top layer of 10 to 30 cm such that the
brighter ground underneath is exposed allowing a better contrast to
the ambient rock (Fig.\ \ref{fig:nazcapath}).
Moisture from morning mist would harden the clay below and shield
the lines from wind preventing erosion.
The conditions of the region, where rainfall averages only 20
minutes a year, facilitated their preservation for centuries.

A more recent dating by Markus Reindel with archaeological means
shows that the region became densely populated between 800 and 200
BCE \cite{reindel:2009}.
It is fairly likely that the precursor of the Nazca people, the
Paracas, influenced the development of the culture which was to
achieve its florescence later on.
About 25 geoglyphs are said to predate the classic Nazca lines.
They exhibit primarily anthropomorphic motifs and are much smaller,
about 10 to 30 m in length, visible from hilltops.

Some researchers allege that similarities would exist on pottery,
but we did not find a fully convincing example to support that
opinion.
The images on pots and bowls are not made of one single line, as
one should expect for a counterpart, but represent rather customary
artwork of the respective culture in ancient Peru.

It can only be guessed how the figures were designed and then
transferred to that large scale.
According to archaeologist Persis Clarkson the construction did not
require complex technology;
all that was needed was the will \cite{takacs-cline}.
Wooden stakes discovered at the end of some lines support the
concept of using very simple equipment.
In this spirit, it may have sufficed to lay a rope on the surface,
and a small number of persons could easily create even the largest
ones within days \cite{nickell:2005}.
The remains of small non-habitational buildings were found in the
immediate vicinity, too \cite{silverman:1990}.
Although best seen from air, the smaller figures can be spotted from
a hill.

Suggestions about the objective have been disputed for many decades,
not to mention the radical hypothesis related to aliens and
supernatural powers.
The history of the more profound research is long, see e.g.\
\cite{vandenbergh:1992, silverman:1990, nickell:2005, takacs-cline,
fathy:2019}.
Starting with the Peruvian archaeologist Toribio Mejia Xesspe
(1896--1983), he was the first to present the objects to the
scientific world in the 1930s.
Unfortunately, not much of his work was reported.
Other names involve Paul Kosok (1896--1959) and Maria Reiche
(1903--1998) who proposed an astronomy-related usage, e.g.\
calendrical pointers, alignments to prominent stars, or earthly
copies of constellations \cite{reiche:1968}.
The archaeoastronomers Gerald Hawkins and Anthony Aveni argued
against an astronomical meaning.
An interesting hypothesis by David Johnson connects to water
(marks to underground springs), which is a very valuable resource
in that region, however, the theory lacked sufficient evidence.
Another approach by Johan Reinhard is that the lines were pathways
for religious ceremonies to be walked upon.
This option is favoured among archaeologists.
But also the idea of a sports complex was raised where runners had
to competitively run around.
Or the figures were sacred assembly areas each used by a different
tribe.
Or they were to represent some kind of ritual mazes.
We relinquish mentioning other suggestions of lesser importance.

%
\fancyhead{}
\fancyhead[C, C]{\footnotesize \itshape \author\ (2021): \titel}
\renewcommand{\headrulewidth}{0pt}


\section{The eclipse hypothesis}

A theory concerning eclipses was disseminated by Robin Edgar on the
Internet \cite{edgar:2000}.
We did not find a solid publication for citation, but this should
not be an argument for rejecting an idea --- it's the thought that
counts!

The lore is based on total solar eclipses that give an impression
of God ``casting an eye'' over what happens on Earth.
According to Edgar, a series of eclipses would have inspired the
Nazca to draw the gigantic pictures to be perceived from above.
He presents a couple of dates whereat he merely hints to the close
succession of some incidents.
Noteworthy are two eclipses within 1.5 years:
3 June -177 and 16 November -178.
Both central tracks did not pass directly over Nazca but in close
distance of 40 and 100 km, respectively.
Two or three more eclipses in the 1st century BCE may also be of
certain interest.

From our point of view, many arguments of Edgar's concept stretch
the interpretation a bit too far.
We cannot support, for example, the parallels drawn by him to
Greek, Assyrian, or Mayan symbols just as little as his comparison
of the totally obscured sun with the pupil in the human's eye to
be ``God's eye in the sky''.
Neither the portrayals on textiles nor other artifacts have provided
convincing evidence regarding the geometric shapes.
Nevertheless, one point prevails worth considering:
eclipses might have given the stimulation.
The figures would function as a medium for communication with the
deities of that people.

As stated in our earlier papers on {\tt arXiv}, a sudden obscuration
of the sun causes panic, however, it is not required to have it
fully covered.
When the sun is about to vanish at an unanticipated time, the
immediate reaction is the anxiety regarding the end of the world.
This fear of destruction is caused by solar eclipses much more than
by earthquakes, hurricanes, floods, or any other natural disaster.
Thus, several striking events in a short time (especially within a
lifetime) would raise the wish for protection against that
``cosmic catastrophe'' to come.

Having such extremely dry conditions, we can assume that almost
every eclipse could have caught attention, even the less conspicuous
ones.
The local circumstances at Nazca turn out different than at most
other places where the ``normal'' weather pattern reduces the number
of successful eclipse sightings.
The only constraint provides the danger of eye damage:
nobody would deliberately stare into the glare of the sun.
Thus, most low-magnitude eclipses would still pass by unperceived,
but a few partial obscurations can justify an accidental notice,
say, down to magnitude $\approx$0.5.
We usually consider mag $>$0.7 as significant.
The sun close to the horizon raises the probability of observation
very much, but the emotional effect will transpire smaller than in
daytime.

\enlargethispage{0.5ex}

\section{Series of high magnitude eclipses}

We examined the list of suitable eclipses for Nazca paying attention
to the historical findings above.
The \textit{Five Millennium Catalog} by Fred Espenak served as an
aid \cite{espenak}.
However, the sheer number of eligible events goes beyond reasoning,
thus, a selection has to be made to special incidents.

There are only four eclipse tracks traversing directly the site of
Nazca during the millennium before its cultural height:
-1091 (total), -372 (annular), -357 (total), and -093 (total).
The eclipse of -357 may turn out pivotal.
An almost ``hit'' occurred on 8 May -555 when an annular eclipse
(mag = 0.960) was progressing at sunset with the maximum taking
place below the line of the ideal horizon.
The central zone of annularity missed Nazca by 45 km.

Comparable to the ``blocks of eclipses'', which we identified for
New Zealand during the Maori era \cite{khalisi-maori}, we 
discovered a number of promising time spans for Nazca, too.
A list for the 1st millennium BCE is given in Table
\ref{tab:nazcablocks} at the end of the paper.
We try to justify our choice of expedient blocks as follows.

In order to make a comparison between eclipses possible, we tried
a very simple approach.
Without taking care of visibility, we limited our attention to
5-year-intervals within which two eclipses of magnitude as little
as $>$0.1 would occur.
A solitary event is of poor interest, and the same could be true
for just two of them in close succession.
We collected eclipses into a block from three to many cases
principally observable from Nazca.
The rupture of the series in the block occurs whenever the next
eclipse lies at least 5 years apart.
This time span seems arbitrary, but it is quite practical and can
be substantiated with the reminiscence of its predecessor.
So, for the 1st millennium BCE, the largest block was found
containing 13 events (it is the very first in our Table).

Then, we used the mean of the magnitudes as an average for that
block.
This procedure only serves as a pre-selection of a series of
potential interest.
It does not imply that the highest block value is automatically
best and bears the result wanted.
It is the trained eye on the data that checks for the essentials
whether or not the circumstances of an eclipse could leave an
impression:
date, time of day, sun height, etc.
Some additional knowledge about the cultural context will certainly
contribute to this purpose.

\begin{figure}[t]
\includegraphics[width=\linewidth]{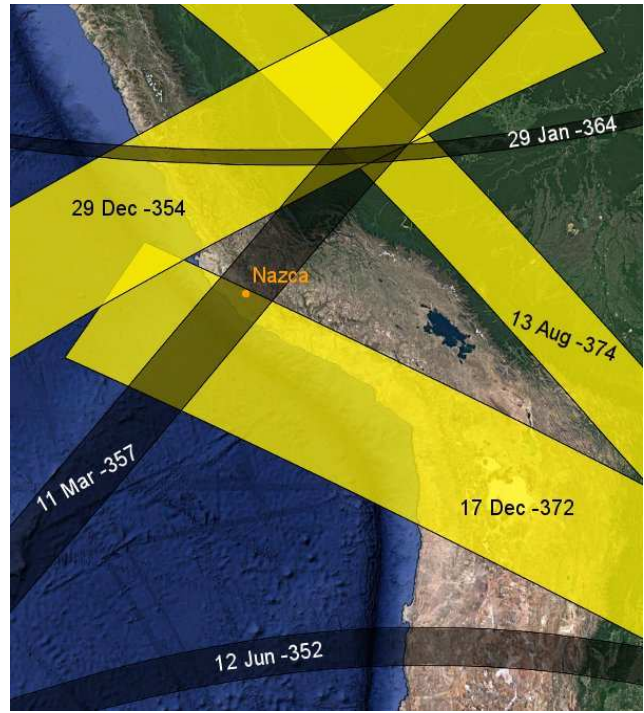}
\caption{Six eclipses of large magnitude within 5 years crossed
    southern Peru (grey = total, yellow = annular).}
\label{fig:peru4thcen}
%
%
\end{figure}

Two blocks, at least, stand out in Table \ref{tab:nazcablocks}.
The most severe reaction could be attributed to the mid-4th century
BCE.
Starting with -374, the next eclipses gradually appeared more
threatening (highlighted in grey in the Table).
At irregular intervals, however, within 3.5 to 1.5 year's time, a
number of large partial obscurations took place with each of them
recalling ``bad memories''.
The climax was reached with the total eclipse in -357 when it
occurred at midday and plunged Nazca into total darkness
(Fig.\ \ref{fig:peru4thcen}).
We think that the chain of events would be sufficient to draw the
necessary consequences for the inhabitants.

\begin{figure}[t]
\includegraphics[width=\linewidth]{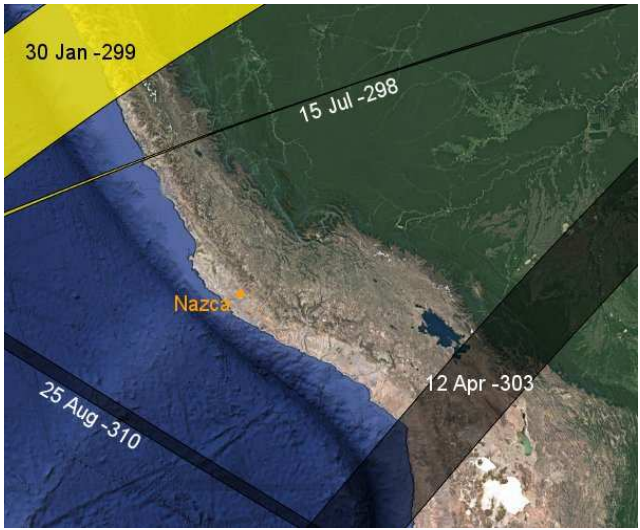}
\caption{Four partial eclipses in Nazca at the turn of the 4th
    century BCE.}
\label{fig:peru3rdcen}
%
%
\end{figure}

Fifty years after that block, another sequence could have raised
attention affecting the next generation
(Fig.\ \ref{fig:peru3rdcen}).
Some elders who experienced the previous series might be still alive
reporting about those strange things in their youth.
Interim eclipses did happen, of course, e.g.\ in -342 (mag = 0.510),
but we disregard them because of their unfavourable viewing
conditions.

Other periods with some potential can be discerned in the 1st
century BCE.
We consider them less valuable for reasons given in the next
section:
maybe they come too late.


\section{Discussion}

The expense to produce these historic geoglyphs on the desert floor
has puzzled generations.
Their most peculiar feature is that they are difficult to see from
the ground.
It seems that they were \emph{intended} to be viewed for a spectator
from above.

Our explanation catches up with the communication hypothesis:
it would not be hard to understand that they were to depict effigies
to be seen by gods.
The metaphor by Robin Edgar may illustrate the idea
\cite{edgar:2000}:
``If you were stranded on a desert island that had a nice stretch
of sandy beach, would you have any difficulty marking out a large
S.O.S.\ signal that was intended to be visible to search and rescue
aircraft?'' ---
The parallel here is that people felt powerless against the various
meteorological and astronomical phenomena, so, they attempted to
please their gods.
Natives are quickly startled when the higher powers up in the sky
``look angry''.
Solar (and lunar) eclipses were known since time immemorial and on
oral transmission they would turn into legends.
Knowing that the sun could fail to shine one day the simple person
will endeavour to construct effigies of living creatures to show
that Earth is inhabited.

The astronomer Sidney van den Bergh pointed out that ancient people
undertook great collective works only if they had a practical
interest (irrigation, fortification),
or because of religious significance \cite{vandenbergh:1992}.
As for the Nazca geoglyphs he gave the religious interpretation the
higher priority.
%
The religious belief of indigenous peoples usually appears
pantheistic and polytheistic, i.e.\ they worshipped nature.
For a civilisation without a written language, as the Nazca was,
it can only be interpreted by deciphering its iconography on
pottery, burials, trophy heads, attire and alike.
Working out the details will be left to the historians, see Helaine
Silverman for examples in the Andes \cite{silverman:2002}.
There, we find a deep devotion to the sun:
Paracas (900--200 BCE), Wari (7th--10th century),
Muisca (15th century), and Inca (13--15th century).
The sun would be part of a pan-Andean religion, as much as it was
recognised the principal deity of almost every culture elsewhere in
the world.

The fear of eclipses runs like a golden thread through the entire
history of mankind and is recorded everywhere over the world
\cite{finsternisbuch}.
Although it is a transient show of light, it concerns the most
important celestial object upon which life depends.
Any native person will instantly be alarmed when the sun is just
going to decease.
The most frequent ``solution'' was making noise to get rid of the
menace.
At Nazca, people could have done so, too, but also created figures
precautionary in the desert sending out signals of life along with
the imploration that the Earth should not be destroyed.
This could be the way how the people communicated with their
pantheon:
spirits of the wind, rain, sky, moon, and the sun, in particular.
What we actually see now is a proliferation of a repetitive activity
that embarked on \emph{before} the Nazca.
This points to a stimulation by the antecessors who experienced
strange happenings, presumably in that 4th century BCE.
Thereupon efforts were magnified, and the works inflated to an
oversize similar to the pyramids to be tombs for the pharaohs.

However, we want to be cautious at interpretation as well.
Since we do not possess written documents, one has to judge the
scientific achievements of a native culture from mathematics, its
architecture, and technological skills.
It is found that the advancements in pre-Columbian South America
were rather low.
Neither the Nazca nor their successors seem to have asked for the
basics of astronomical phenomena.
Except the winter solstice we do not meet any knowledge about the
ecliptic, equator, the path of the moon, periodic visibility of
planets, conjunctions, and even less the basics behind eclipses.
The scientific issues on the whole continent seem not as progressive
as the Mayan in Central Mexico or in the Old World (Mesopotamia,
Egypt, China, India).
None of the civilisations developed the mathematics needed to
follow a cycle.
Such level is essential not just for time reckoning, but for any
kind of celestial observation (compare this with the sophisticated
Mayan calendar).
Mathematical knowledge often manifests itself in many, many other
ideas leading to advanced technology \cite{khalisi-progress},
but no such technology emerged.
Natives took the heavenly incidents as they occurred.

It may be instructive to remember that astronomy in ancient China
did not rise above a rudimentary level, either.
The royal astronomers of former times were careful observers and
they did write down any small changes in the sky like comets,
supernovae and meteorological appearances leaving behind a wonderful
sample of precious data.
However, they did not elaborate a ``cosmology'' out of their
observations.
The theory of planetary movements and models of the universe are
the merit of the Babylonians and of the Greeks.
It was them who tried to find a mechanism in order to make
predictions for future incidents, but not the Chinese.
In India, too, science ignited after Alexander's campaign in the
4th century BCE bringing along Greek wisdom.

The author of this paper is an astrophysicist, not an archaeologist.
He does not insist on the explanation given here.
Nevertheless, he is convinced that the thought about eclipses as
the basic cause for the geoglyphs is worth considering because of
plausibility.
He thinks that one should not put too much science into these
geoglyphs.
We should look upon them as what they are:
effigies for a conceived deity in the sky.
If this explanation should turn out false, either, we hope to pave
the way for further suggestions.


\section*{Acknowledgements}

This paper is based on Chapter 8.2 of the Habilitation submitted
to the University of Heidelberg one year ago (Feb 2020).
Here we present an improved and updated work.

During the entire project,
the author had to cope with repression by fraudulent judiciary
and immense brutality by officials.
All attacks were to deprive him of occupational advancement.
Details will be published in an autobiography accessible via the
German National Library.

The author is very, very grateful to his family members as well as
his long-time friends for their support against the intrigues.


\begin{table*}[t]
\caption{Solar eclipses at Nazca.
    A $\ast$-star in the ``Type'' column (total, annular, hybrid)
    indicates that Nazca was in the central path.
    ``$\odot$-h.'' denotes the height of the sun at the time of
    maximum.
    The last column gives the average magnitude for the block.
    The prominent cases of Figures \ref{fig:peru4thcen} and
    \ref{fig:peru3rdcen} are highlighted.}
\label{tab:nazcablocks}
\vspace{-\baselineskip}
\changefontsizes{9pt}
\begin{multicols}{2}
\centering
\begin{tabular}{lccrr|c}
%
\hline
\rowcolor{grey20}
   Date     &Type&   LT  & Magn. & $\odot$-h. & $\varnothing$ block \\ 
\hline
-979 Aug 13 &  T & 16:33 & 0.918 &  15.6 & \multirow{13}{*}{0.528} \\
-978 Feb 05 &  A & 06:59 & 0.170 &  14.2 &  \\
-976 Jun 10 &  A & 13:04 & 0.743 &  48.9 &  \\
-976 Dec 05 &  A & 16:23 & 0.898 &  26.2 &  \\
-972 Mar 30 &  A & 09:56 & 0.357 &  54.5 &  \\
-971 Sep 12 &  T & 08:22 & 0.441 &  31.7 &  \\
-967 Jan 05 &  T & 13:54 & 0.144 &  63.7 &  \\
-965 Nov 04 &  A & 12:49 & 0.313 &  75.1 &  \\
-964 Apr 30 &  T & 07:12 & 0.705 &  15.2 &  \\
-964 Oct 23 &  A & 17:46 & 0.429 &   2.9 &  \\
-962 Sep 03 &  T & 07:31 & 0.207 &  18.3 &  \\
-961 Feb 23 &  A & 08:07 & 0.576 &  29.4 &  \\
-960 Feb 16 &  H & 18:28 & 0.960 &   1.5 &  \\
\hline
-925 Sep 14 &  A & 16:16 & 0.413 &  22.4 & \multirow{5}{*}{0.539} \\
-924 Mar 09 &  T & 06:40 & 0.255 &   8.3 &  \\
-921 Jan 07 &  A & 17:46 & 0.289 &  11.0 &  \\
-917 Apr 21 &  A & 07:40 & 0.806 &  22.0 &  \\
-917 Oct 15 &  T & 10:52 & 0.933 &  72.8 &  \\
\hline
-885 Jan 25 &  A & 05:38 & 0.364 &   9.9 & \multirow{7}{*}{0.489} \\
-885 Jul 24 &  T & 13:58 & 0.632 &  43.1 &  \\
-882 Nov 17 &  T & 14:03 & 0.690 &  57.6 &  \\
-878 Mar 12 &  T & 15:48 & 0.647 &  34.4 &  \\
-877 Aug 24 &  A & 10:17 & 0.362 &  50.9 &  \\
-874 Dec 17 &  A & 15:02 & 0.566 &  61.2 &  \\
-871 Oct 16 &  A & 17:20 & 0.164 &   8.9 &  \\
\hline
-860 Sep 15 &  A & 16:32 & 0.371 &  18.7 & \multirow{4}{*}{0.449} \\
-856 Jan 08 &  A & 14:50 & 0.375 &  51.3 &  \\
-855 Jun 23 &  A & 14:28 & 0.880 &  36.0 &  \\
-853 Apr 23 &  T & 15:42 & 0.172 &  29.4 &  \\
\hline
-845 Jun 03 &  A & 09:43 & 0.546 &  42.5 & \multirow{11}{*}{0.586} \\
-835 Nov 07 &  A & 06:50 & 0.942 &  18.3 &  \\
-834 Oct 27 &  A & 12:48 & 0.457 &  74.7 &  \\
-833 Apr 22 &  T & 15:02 & 0.593 &  38.3 &  \\
-831 Aug 25 &  T & 13:41 & 0.794 &  52.1 &  \\
-830 Feb 19 &  A & 07:44 & 0.254 &  24.1 &  \\
-828 Dec 19 &  T & 17:35 & 0.255 &  11.5 &  \\
-827 Dec 08 &  T & 06:09 & 0.721 &   8.7 &  \\
-824 Apr 12 &  T & 15:14 & 0.847 &  37.8 &  \\
-823 Sep 25 &  A & 09:56 & 0.376 &  56.0 &  \\
-819 Jan 18 &  H & 12:12 & 0.663 &  82.8 &  \\
\hline
-812 Mar 02 &  A & 18:29 & 0.620 &  -0.4 & \multirow{6}{*}{0.430} \\
-812 Aug 26 &  T & 14:56 & 0.148 &  38.2 &  \\
-809 Dec 20 &  T & 18:19 & 0.637 &   1.8 &  \\
-806 Oct 18 &  A & 15:46 & 0.353 &  31.5 &  \\
-802 Feb 10 &  A & 13:52 & 0.373 &  67.3 &  \\
-801 Jul 26 &  A & 08:46 & 0.448 &  29.9 &  \\
\hline
-781 Dec 10 &  A & 08:24 & 0.298 &  29.6 & \multirow{3}{*}{0.659} \\
-780 Nov 28 &  A & 13:15 & 0.702 &  69.3 &  \\
-779 May 24 &  T & 12:57 & 0.976 &  52.2 &  \\
\hline
-773 Jul 16 &  A & 14:22 & 0.839 &  38.4 & \multirow{5}{*}{0.707} \\
-772 Jan 10 &  T & 08:35 & 0.498 &  38.7 &  \\
-770 May 15 &  T & 13:28 & 0.769 &  50.4 &  \\
-769 Oct 28 &  A & 11:11 & 0.964 &  80.2 &  \\
-765 Feb 20 &  T & 12:39 & 0.467 &  84.7 &  \\
\hline
-758 Apr 04 &  A & 14:04 & 0.438 &  55.8 & \multirow{3}{*}{0.527} \\
-758 Sep 27 &  T & 16:14 & 0.826 &  23.6 &  \\
-757 Mar 24 &  A & 16:08 & 0.316 &  29.8 &  \\
\hline
%
\end{tabular}


\centering
\begin{tabular}{lccrr|c}
\hline
\rowcolor{grey20}
   Date     &Type&   LT  & Magn. & $\odot$-h. & $\varnothing$ block \\
\hline
-752 Nov 18 &  A & 16:13 & 0.304 &  27.0 & \multirow{3}{*}{0.413} \\
-751 May 15 &  T & 11:26 & 0.300 &  57.7 &  \\
-750 Oct 28 &  T & 07:00 & 0.635 &  20.1 &  \\
\hline
-741 Oct 19 &  T & 05:56 & 0.928 &   3.6 & \multirow{5}{*}{0.573} \\
-737 Feb 11 &  T & 10:35 & 0.321 &  65.3 &  \\
-729 Sep 07 &  A & 16:33 & 0.470 &  18.0 &  \\
-726 Dec 31 &  A & 13:25 & 0.684 &  69.3 &  \\
-725 Jun 26 &  T & 10:44 & 0.462 &  47.9 &  \\
\hline
-719 Aug 17 &  A & 08:48 & 0.822 &  33.0 & \multirow{6}{*}{0.474} \\
-718 Feb 11 &  T & 10:50 & 0.447 &  68.7 &  \\
-718 Aug 07 &  A & 16:10 & 0.312 &  20.2 &  \\
-716 Jun 16 &  T & 11:13 & 0.221 &  50.8 &  \\
-715 Nov 29 &  A & 14:10 & 0.413 &  56.6 &  \\
-711 Mar 24 &  T & 11:31 & 0.629 &  73.7 &  \\
\hline
-676 Apr 26 &  A & 17:01 & 0.683 &  10.3 & \multirow{9}{*}{0.766} \\
-675 Oct 10 &  A & 15:21 & 0.543 &  37.2 &  \\
-671 Feb 01 &  A & 11:57 & 0.809 &  83.6 &  \\
-668 Nov 20 &  T & 09:46 & 0.731 &  60.2 &  \\
-665 Sep 19 &  A & 06:43 & 0.582 &  10.1 &  \\
-664 Mar 15 &  T & 10:46 & 0.959 &  66.6 &  \\
-664 Sep 07 &  A & 14:51 & 0.740 &  41.2 &  \\
-660 Jan 01 &  H & 16:53 & 0.905 &  22.4 &  \\
-657 Apr 26 &  T & 09:25 & 0.938 &  44.7 &  \\
\hline
-625 Jul 31 &  H & 16:21 & 0.407 &  17.1 & \multirow{12}{*}{0.664} \\
-624 Jan 23 &  H & 06:25 & 0.902 &   7.4 &  \\
-621 Nov 12 &  A & 15:25 & 0.529 &  37.7 &  \\
-617 Mar 06 &  A & 08:30 & 0.669 &  34.9 &  \\
-614 Dec 23 &  T & 15:02 & 0.977 &  60.4 &  \\
-611 Oct 21 &  A & 06:21 & 0.406 &  10.2 &  \\
-610 Apr 17 &  T & 08:39 & 0.635 &  35.6 &  \\
-610 Oct 10 &  A & 15:31 & 0.767 &  34.7 &  \\
-606 Feb 03 &  T & 18:01 & 0.403 &   8.4 &  \\
-603 May 28 &  T & 07:14 & 0.444 &  13.5 &  \\
-600 Mar 27 &  A & 17:29 & 0.879 &   9.0 &  \\
-598 Aug 30 &  H & 06:36 & 0.949 &   5.4 &  \\
\hline
-587 Feb 02 &  T & 17:21 & 0.257 &  17.6 & \multirow{3}{*}{0.460} \\
-585 Jun 09 & (T)& 17:45 & 0.925 &  -4.0 &  \\
-584 May 28 &  T & 07:35 & 0.199 &  18.0 &  \\
\hline
-578 Jan 25 &  T & 16:46 & 0.560 &  25.8 & \multirow{9}{*}{0.593} \\
-570 Feb 24 &  T & 07:01 & 0.949 &  13.6 &  \\
-567 Dec 13 &  A & 15:57 & 0.730 &  33.1 &  \\
-563 Oct 01 &  T & 11:46 & 0.390 &  76.0 &  \\
-562 Mar 27 &  A & 06:54 & 0.348 &  11.5 &  \\
-559 Jan 24 &  T & 16:17 & 0.279 &  32.4 &  \\
-556 May 16 &  T & 05:58 & 0.262 &  -3.1 &  \\
-556 Nov 11 &  A & 17:34 & 0.855 &   7.3 &  \\
-555 May 08 &  A & 17:38 & 0.960 &  -0.4 &  \\
\hline
-534 Sep 11 &  A & 06:47 & 0.245 &  10.0 & \multirow{4}{*}{0.549} \\
-533 Mar 07 &  T & 18:00 & 0.838 &   5.2 &  \\
-531 Jul 11 &  T & 16:09 & 0.603 &  17.9 &  \\
-530 Jun 30 &  T & 06:07 & 0.510 &  -3.4 &  \\
\hline
-524 Feb 26 &  T & 17:46 & 0.892 &  10.1 & \multirow{8}{*}{0.753} \\
-523 Feb 15 &  T & 05:54 & 0.721 &  -1.9 &  \\
-520 Dec 03 &  H & 07:41 & 0.932 &  30.2 &  \\
-516 Mar 28 &  T & 06:30 & 0.822 &   5.7 &  \\
-516 Sep 21 &  A & 17:56 & 0.614 &  -1.0 &  \\
-512 Jan 15 &  A & 15:31 & 0.850 &  42.5 &  \\
-512 Jul 11 &  T & 17:14 & 0.340 &   4.1 &  \\
-509 Nov 03 &  T & 14:33 & 0.855 &  49.9 &  \\
\hline
\end{tabular}
\end{multicols}
\end{table*}


\begin{table*}[t]
\caption*{Table 1: Continued.}
\vspace{-\baselineskip}
\changefontsizes{9pt}
\begin{multicols}{2}
\centering
\begin{tabular}{lccrr|c}
\hline
\rowcolor{grey20}
   Date     &Type&   LT  & Magn. & $\odot$-h. & $\varnothing$ block \\
\hline
-465 Jan 05 &  H & 10:45 & 0.459 &  68.6 & \multirow{8}{*}{0.507} \\
-462 Oct 25 &  A & 17:23 & 0.539 &   8.5 &  \\
-461 Oct 14 &  A & 16:35 & 0.260 &  19.3 &  \\
-458 Feb 16 &  A & 13:10 & 0.634 &  77.2 &  \\
-458 Aug 13 &  T & 16:28 & 0.955 &  17.1 &  \\
-457 Feb 05 &  A & 15:25 & 0.224 &  45.0 &  \\
-455 Dec 05 &  T & 17:45 & 0.816 &   7.5 &  \\
-451 Mar 31 &  T & 15:59 & 0.167 &  29.9 &  \\
\hline
-436 Jun 10 &  H & 07:28 & 0.563 &  15.2 & \multirow{5}{*}{0.499} \\
-431 Sep 12 &  P & 06:14 & 0.234 &   2.4 &  \\
-429 Jul 23 &  A & 16:21 & 0.563 &  16.7 &  \\
-426 Nov 15 &  A & 05:43 & 0.629 &   3.4 &  \\
-425 May 12 &  T & 16:32 & 0.504 &  14.1 &  \\
\hline
-419 Jul 02 &  A & 09:46 & 0.439 &  39.9 & \multirow{3}{*}{0.443} \\
-419 Dec 27 &  H & 08:17 & 0.582 &  36.0 &  \\
-416 May 02 &  P & 16:35 & 0.307 &  15.0 &  \\
\hline
-411 Feb 06 &  T & 13:09 & 0.797 &  77.5 & \multirow{7}{*}{0.506} \\
-408 Nov 26 &  A & 17:33 & 0.730 &   9.1 &  \\
-407 Nov 14 &  A & 16:52 & 0.204 &  17.4 &  \\
-406 May 11 &  T & 15:09 & 0.285 &  32.1 &  \\
-404 Mar 20 &  A & 08:34 & 0.519 &  35.5 &  \\
-404 Sep 14 &  T & 16:51 & 0.499 &  14.3 &  \\
-403 Mar 09 &  A & 13:07 & 0.510 &  74.2 &  \\
\hline
\cellcolor{grey10} -374 Aug 13 &  A & 11:29 & 0.821 &  57.5 & \multirow{10}{*}{0.759} \\
\cellcolor{grey10} -372 Dec 17 & A* & 05:47 & 0.908 &   3.2 &  \\
-368 Oct 05 &  T & 06:36 & 0.608 &  11.9 &  \\
\cellcolor{grey10} -364 Jan 29 &  H & 10:35 & 0.888 &  65.4 &  \\
-364 Jul 23 &  H & 15:27 & 0.717 &  27.9 &  \\
-361 Nov 16 &  H & 12:03 & 0.531 &  85.3 &  \\
-360 May 12 &  A & 07:50 & 0.534 &  22.4 &  \\
\cellcolor{grey10} -357 Mar 11 & T* & 13:19 & 1.020 &  71.1 &  \\
\cellcolor{grey10} -354 Dec 29 &  A & 17:35 & 0.866 &  12.6 &  \\
\cellcolor{grey10} -352 Jun 12 &  T & 11:24 & 0.781 &  51.5 &  \\
-349 Apr 11 &  A & 09:00 & 0.680 &  40.6 &  \\
\hline
-320 Sep 14 &  A & 08:20 & 0.372 &  32.8 & \multirow{11}{*}{0.635} \\
-316 Jan 08 &  A & 11:40 & 0.268 &  12.7 &  \\
-314 Nov 07 &  H & 08:51 & 0.760 &  47.6 &  \\
-312 Apr 21 &  T & 12:44 & 0.205 &  62.2 &  \\
-310 Mar 02 &  H & 11:00 & 0.464 &  70.8 &  \\
\cellcolor{grey10} -310 Aug 25 &  H & 13:38 & 0.869 &  53.5 &  \\
-307 Dec 18 &  H & 15:31 & 0.617 &  39.7 &  \\
\cellcolor{grey10} -303 Apr 12 &  T & 12:09 & 0.830 &  67.9 &  \\
\cellcolor{grey10} -299 Jan 30 &  A & 16:43 & 0.765 &  26.5 &  \\
\cellcolor{grey10} -298 Jul 15 &  H & 07:43 & 0.874 &  16.8 &  \\
-295 May 13 &  H & 05:35 & 0.957 &  -8.2 &  \\
\hline
-288 Jun 24 &  A & 11:08 & 0.167 &  49.8 & \multirow{11}{*}{0.448} \\
-277 Nov 18 &  A & 13:19 & 0.518 &  67.9 &  \\
-273 Mar 13 &  A & 08:56 & 0.241 &  40.9 &  \\
-266 Oct 16 &  A & 07:35 & 0.355 &  27.8 &  \\
-262 Feb 09 &  A & 06:32 & 0.674 &   7.5 &  \\
-260 Dec 09 &  H & 12:51 & 0.127 &  74.8 &  \\
-258 May 24 &  T & 09:49 & 0.629 &  44.0 &  \\
-256 Sep 26 &  H & 13:50 & 0.260 &  57.4 &  \\
-251 Jan 09 &  T & 05:56 & 0.818 &   2.3 &  \\
-249 May 15 &  T & 10:03 & 0.331 &  47.9 &  \\
-245 Mar 03 &  A & 15:11 & 0.807 &  60.8 &  \\
\hline
-237 Sep 27 &  T & 12:37 & 0.275 &  71.5 &  \\ 
-234 Jul 27 &  A & 05:52 & 0.900 &  -7.3 &  \\
-233 Jan 21 &  T & 17:57 & 0.980 &   9.2 &  \\
    \multicolumn{5}{c|}{\dots }          &  \\
\end{tabular}


\centering
\begin{tabular}{lccrr|c}
\hline
\rowcolor{grey20}
   Date     &Type&   LT  & Magn. & $\odot$-h. & $\varnothing$ block \\
\hline
    \multicolumn{5}{c|}{\dots }          &  \\
-230 May 15 &  T & 10:38 & 0.232 &  52.6 &      0.540 \\
-229 Oct 28 &  A & 06:27 & 0.502 &  12.7 &  \\
-226 Aug 27 &  A & 16:21 & 0.289 &  20.1 &  \\
-223 Dec 20 &  A & 14:53 & 0.335 &  48.7 &  \\
\hline
-216 Feb 11 &  T & 07:29 & 0.357 &  20.8 & \multirow{3}{*}{0.550} \\
-216 Aug 06 &  A & 16:14 & 0.593 &  19.8 &  \\
-212 Nov 18 &  A & 08:00 & 0.700 &  35.4 &  \\
\hline
-205 Jan 11 &  H & 15:07 & 0.702 &  38.5 & \multirow{5}{*}{0.528} \\
-204 Jun 25 &  T & 07:31 & 0.951 &  14.7 &  \\
-201 Apr 25 &  A & 15:12 & 0.222 &  34.5 &  \\
-200 Apr 13 &  A & 16:06 & 0.394 &  24.6 &  \\
-197 Feb 11 &  T & 07:50 & 0.372 &  25.6 &  \\
\hline
-166 Nov 20 &  T & 06:24 & 0.627 &  13.1 & \multirow{12}{*}{0.426} \\
-162 Mar 15 &  T & 07:49 & 0.166 &  24.6 &  \\
-162 Sep 08 &  A & 11:48 & 0.556 &  67.9 &  \\
-161 Aug 28 &  A & 13:08 & 0.128 &  59.0 &  \\
-158 Dec 21 &  A & 08:56 & 0.731 &  45.6 &  \\
-155 Oct 20 &  H & 13:57 & 0.233 &  57.8 &  \\
-151 Feb 12 &  H & 16:38 & 0.860 &  27.2 &  \\
-150 Jul 28 &  T & 06:20 & 0.392 &  -1.1 &  \\
-147 May 27 &  A & 07:42 & 0.196 &  19.2 &  \\
-147 Nov 20 &  T & 06:29 & 0.121 &  14.1 &  \\
-146 May 16 &  A & 08:56 & 0.450 &  35.5 &  \\
-143 Mar 15 &  T & 08:43 & 0.648 &  37.6 &  \\
\hline
-136 Apr 25 &  H & 11:28 & 0.380 &  62.1 & \multirow{4}{*}{0.504} \\
-135 Oct 09 &  A & 06:37 & 0.611 &  13.0 &  \\
-132 Aug 07 &  T & 17:19 & 0.540 &   5.6 &  \\
-131 Jul 28 &  T & 07:17 & 0.485 &  11.7 &  \\
\hline
-122 Jul 19 &  T & 06:50 & 0.929 &   5.5 & \multirow{4}{*}{0.455} \\
-119 May 17 &  A & 13:21 & 0.448 &  49.9 &  \\
-118 Oct 31 &  A & 16:13 & 0.323 &  25.6 &  \\
-114 Feb 23 &  A & 13:34 & 0.120 &  70.4 &  \\
\hline
-108 Oct 10 &  A & 09:12 & 0.367 &  50.6 & \multirow{8}{*}{0.609} \\ 
-107 Sep 29 &  A & 10:42 & 0.385 &  68.2 &  \\
-106 Mar 27 &  T & 17:57 & 0.481 &   1.9 &  \\
-103 Jan 22 &  A & 09:58 & 0.530 &  57.0 &  \\
-101 Nov 22 &  H & 16:24 & 0.287 &  24.5 &  \\
-097 Mar 17 &  H & 15:54 & 0.923 &  33.3 &  \\
-093 Dec 23 & T* & 09:53 & 1.019 &  58.4 &  \\ 
-089 Apr 17 &  T & 07:54 & 0.881 &  25.0 &  \\
\hline
-082 May 28 &  T & 07:50 & 0.770 &  20.6 & \multirow{6}{*}{0.568} \\
-082 Nov 22 &  A & 05:40 & 0.236 &   2.9 &  \\
-081 Nov 11 &  A & 06:07 & 0.524 &   9.1 &  \\
-078 Sep 10 &  T & 17:34 & 0.514 &   3.7 &  \\
-077 Aug 30 &  T & 07:00 & 0.901 &  11.5 &  \\
-074 Jun 28 &  A & 12:57 & 0.462 &  48.8 &  \\
\hline
-068 Aug 20 &  T & 06:02 & 0.781 &  -3.3 & \multirow{3}{*}{0.766} \\
-064 Dec 02 &  A & 17:24 & 0.652 &  12.0 &  \\
-060 Mar 27 &  A & 09:52 & 0.866 &  53.0 &  \\
\hline
-053 Nov 01 &  A & 10:36 & 0.862 &  73.2 & \multirow{3}{*}{0.825} \\
-052 Apr 27 &  T & 15:56 & 0.841 &  24.1 &  \\
-049 Feb 24 &  A & 09:38 & 0.771 &  51.3 &  \\
\hline
-043 Apr 18 &  A & 13:14 & 0.548 &  58.7 & \multirow{4}{*}{0.474} \\
-038 Jan 24 &  T & 13:34 & 0.707 &  70.6 &  \\
-037 Jul 09 &  H & 06:40 & 0.268 &   3.1 &  \\
-035 May 29 &  T & 06:13 & 0.371 &   0.0 &  \\
\hline
-014 Sep 22 &  T & 06:11 & 0.691 &   4.1 & \multirow{4}{*}{0.665} \\
-012 Mar 06 &  T & 15:26 & 0.414 &  41.9 &  \\
-010 Jul 10 &  T & 15:31 & 0.785 &  26.0 &  \\
-009 Jan 05 &  A & 18:01 & 0.772 &   7.5 &  \\
\hline
%
\end{tabular}
\end{multicols}
\end{table*}


\vspace{\baselineskip}

\bibliography{v192-biblio}

\end{document}